# Implementation of a 12-Million Hodgkin-Huxley Neuron Network on a Single Chip

Byungik Ahn
Neurocoms Inc.
Seoul, South Korea
jerryahn@neurocoms.com

## ABSTRACT
Understanding the human brain is the biggest challenge for scientists in the twenty-first century. The Hodgkin-Huxley (HH) model is one of the most successful mathematical models for bio-realistic simulations of the brain. However, the simulation of HH neurons involves complex computation, which makes the implementation of large-scale brain networks difficult. In this paper, we propose a hardware architecture that efficiently computes a large-scale network of HH neurons. This architecture is based on the neuron machine hardware architecture, which has the limitation of speed as it has only one computation node. The proposed architecture is essentially a non-Von Neumann synchronous system with multiple computation nodes, called hardware neurons, to achieve linear speedup. In this paper, the design of a digital circuit that computes large-scale networks of HH neurons is presented as an example to provide a detailed description of the proposed architecture. This design supports axonal conduction delay of spikes and short- and long-term plasticity synapses, along with floating-point precision HH neurons. The design is implemented on a field-programmable gate array (FPGA) chip and computes a network of one million HH neurons in near real time. The implemented system can compute a network with up to 12 million HH neurons and 600 million synapses. The proposed design method can facilitate the design of systems supporting complex neuron models and their flexible implementation on reconfigurable FPGA chips.

## 1 Introduction

Research investigations of an artificial brain are important to understand the underlying causes of brain diseases [1]. The artificial brains can also be used in neuro-prosthetic devices that replace human visual or auditory sensors [2]. As deep learning technology was made possible by a large amount of data and high-performance computing [3], experiments with large-scale artificial brains using high-speed computing devices might lead to the development of new machine learning algorithms [4]. In addition, based on the design of computers that run artificial brains efficiently, a small, low-power computer with high efficiency, which is completely different from a conventional computer can be designed. Owing to the abovementioned reasons, major projects such as the Human Brain Project in Europe [1] and the Brain Initiative in the United States [5] have been launched worldwide.

The Hodgkin-Huxley (HH) model is the most fundamental and widely known neuron model among many neural computational models. The HH is bio-realistic as it reproduces the neuronal mechanism with the flow of ion currents, and the model can be extended to reflect new ion currents. However, the HH model involves complex computation. For example, the integrate-and-fire model requires 5 FLOPS per 1 ms model time for a neuron, whereas the HH model requires 1200 FLOPS to compute the same time period [6]. Owing to this problem, there are little known large-scale HH systems and the computation speed of the HH system is slow if any. For example, in [7], the original HH model consisting of 3 ion currents was extended to 11 ion currents, and a 400,000-neuron network was computed on a high-end GPU at a speed of 14400 s per 1 s model time. To address this problem, simple phenomenological models such as the Izhikevich model are often used to implement large-scale brain networks. However, these models do not accurately describe all the physical mechanisms of neurons [7] and are difficult to incorporate new phenomena observed in the biological neurons.

In this study, a hardware architecture that can efficiently compute a large network of computationally complex model neurons such as the HH model is proposed. This architecture is based on the neuron machine (NM) architecture, which has been applied to deep learning algorithms such as the back-propagation algorithm [8], deep belief network [9], and convolutional neural networks [10]. It has also been applied to neuromorphic models such as Izhikevich [11] and HH [12] models and used to design and implement small-scale systems computing 1000 neurons. However, the NM architecture describes only one computation node in principle, these systems do not have a computation speed higher than one neuron per clock cycle.

The hardware architecture proposed in this paper is non-Von-Neumann and synchronous, wherein the entire system is designed as a digital circuit operated by a single system clock. In addition, all units in the system are designed as fully pipelined circuits, where data are processed in a time-division manner in a predetermined sequence, with all arithmetic operators in the circuit utilized at a rate of nearly 100% during the entire computation period to achieve high computational efficiency. The most notable feature of the proposed architecture is that it can comprise multiple computation nodes called hardware neurons (HNs) in a system, achieving linear speedup. Furthermore, this architecture imposes no restrictions on the network topology.

This paper presents the design of a system that runs large-scale networks of HH neurons using multiple HNs to provide a detailed description of the proposed architecture. This design supports the HH neuron model as well as axonal conduction delay and synapses with plasticity features. The design was implemented on a single field-programmable gate array (FPGA) chip. The implemented system can run a network of 1 million neurons in near real time and a network of 12 million neurons and 600 million synapses at a speed of 125 s per 1 s model time.

The remainder of this paper is structured as follows. Section 2 describes model neuron for our design and fully pipelined circuit design, while Section 3 presents a hardware design of the proposed hardware architecture. Section 4 evaluates the implementation results. Finally, Sections 5 and 6 present the discussion and the conclusions, respectively.

## 2 Background

### 2.1 Model Neuron

In this section, the computational model of the neurons supported by the systems designed and implemented in this paper is described, as shown in Figure 1.

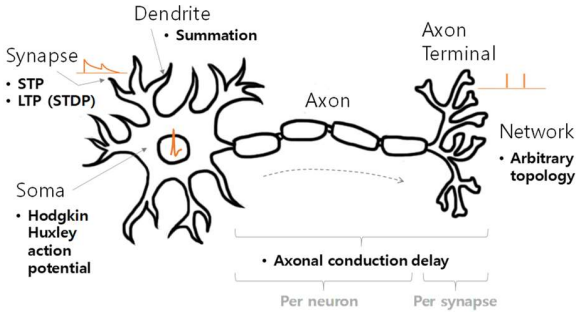

**Figure 1: Features of model neuron**

The brain consists of a network of neurons that send and receive electronic spike signals. When a neuron receives a spike signal from another neuron, a synapse sends a current that changes over time to the cell body of the neuron. The dendrites sum up these current signals, and the soma, which is the body of the neuron, creates an action potential in response to the input current. The spike induced from this action potential is transmitted to other neurons through the axon terminal.

**Short-term plasticity (STP)**. The STP, whose conductance pattern changes according to the history of the recent input spikes, uses the phenomenological model proposed by [13], as

$$\frac{du}{dt} = -\frac{u}{\tau_f} + U(1 - u^-)\delta(t - t_{sp}), \quad (1)$$
$$\frac{dx}{dt} = \frac{1-x}{\tau_d} + u^+ x^- \delta(t - t_{sp}),$$
$$\frac{dS}{dt} = -\frac{S}{\tau_s} + Au^+ x^- \delta(t - t_{sp}).$$

Here, $u, x, S$ are state variables and $U, A, \tau_f, \tau_d, \tau_s$ are attributes specific to the synapse.

**Long-term plasticity (LTP)**. The LTP, which performs a function corresponding to the long-term memory of the brain, is computed using the spike-timing-dependent plasticity (STDP) model [14], as

$$\frac{dx_j}{dt} = -\frac{x_j}{\tau_+} + a_+(x_j) \sum_f \delta\left(t - t_j^f\right), \quad (2)$$
$$\frac{dy}{dt} = -\frac{y}{\tau_-} + a_-(y) \sum_n \delta(t - t^n),$$
$$\frac{dw_j}{dt} = A_+(w_j)x(t)\sum_n \delta(t - t^n) - A_-(w_j)y(t)\sum_f \delta\left(t - t_j^f\right),$$
$$A_+(w_j) = (w^{max} - w_j)\eta_+ \text{ and } A_-(w_j) = w_j\eta_-.$$

Here, $x$ and $w$ are state variables and $a_+, a_-, \tau_+, \tau_-, w^{max}, \eta_+, \eta_-$ are attributes specific to the synapse. In addition, $y$ is a state variable specific to the postsynaptic neuron. In STDP, the synaptic weight, $w$, changes in an unsupervised learning manner based on the spike timing of the presynaptic and postsynaptic neurons.

**Synaptic membrane**. Finally, the synaptic currents can be computed as follows.

$$I_{syn} = S \times w \times g_{syn} \times (E_{syn} - V_{post}) \quad (3)$$

Here, $g_{syn}$ and $E_{syn}$ are synaptic attributes, and $V_{post}$ is the action potential of the postsynaptic neuron.

**Dendrite**. The current values of synapses are all summed and sent to the neuron body.

**Hodgkin-Huxley action potential**. The membrane potential $V$ of a neuron is represented with respect to time by

$$C_m \frac{dV(t)}{dt} = I_{ext}(t) - \sum_i I_i(t, V) \quad (4)$$

where $C_m$ denotes the capacitance of the neuron body, and $I_{ext}$ represents an input current. Each ion current $I_i$ is given by Ohm's law as

$$I(t, V) = g(t, V) \cdot (V - V_{eq}) \quad (5)$$

where $V_{eq}$ denotes the equilibrium potential of the ion and $g(t,V)$ represents the conductance, which can be expanded in terms of its constant average $\bar{g}$ and the activation and inactivation fractions $m$ and $h$, respectively, that determine how much of ions can flow through the available membrane channels. This expansion is given by

$$g(t, V) = \bar{g} \cdot m(t, V)^p \cdot h(t, V)^q \quad (6)$$

and the fractions follow the first-order kinetics

$$\frac{dm(t,V)}{dt} = \frac{m_\infty(V) - m(t,V)}{\tau_m(V)} = \alpha_m(V) \cdot (1 - m) - \beta_m(V) \cdot m \quad (7)$$

with similar dynamics for $h$, where $\alpha$ and $\beta$ denote gate functions, and $m_\infty$ and $\tau_m$ represent steady-state activation and the time constant, respectively.

Based on the above form, a number of individual ion currents can be included. In the proposed design, an inward Na+ current and an outward K+ current are included along with a leak current. Detailed equations of the gate functions for each ion currents can be found in [15].

**Axonal conduction delay**. Actional conduction delay is supported by per-neuron delay applied equally to all output connections of each neuron and per-synapse delay specified differently for each output connection.

**Network**. Any neuron can connect to any other neuron, with no limit to the number of inbound and outbound connections.

## 2.2 Fully Pipelined Circuit

All the units constituting the proposed system are designed as fully pipelined circuits. As an example of the fully pipelined circuit, a simple circuit that computes the conductance of $N$ synapses is considered using the following formula.

$$x_i(T+1) = \begin{cases} 1, & spike_i = 1 \\ x_i(T) - \frac{x_i(T)}{\tau}, & spike_i = 0 \end{cases} \quad (8)$$

The input of this function is the spike information of the presynaptic neuron, while the output is the synaptic conductance. The conductance changes to a peak value of 1.0 each time the spike of the presynaptic neuron is transmitted and decays with the time constant τ. The above equation can be computed by implementing the electronic circuit shown in Figure 2.

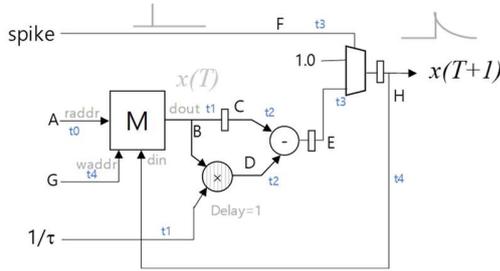

**Figure 2: Example pipelined circuit**

In the dual-port memory M, which can be read and written simultaneously, the conductance values of all $N$ synapses are stored, where the conductance of the $i^{th}$ synapse is stored at the $i^{th}$ address. As the address value of signal point A increases from 0 to $N$ - 1 by 1 every clock cycle via the control circuit (CU), the circuit computes Equation 8 sequentially from synapse 0 to synapse $N-1$. After the computation of synaptic conductance at H, the conductance is written back in the M memory at the next clock cycle. It should be noted that the time reference specified by *tnn* in Figure 2 indicates the time difference relative to t0. Therefore, the computation is performed step by step as the clock cycle progresses. At the next clock cycle of each signal point, the next synapse value is placed and processed in the same sequence. Such an electronic circuit can be designed using a hardware description language. For example, the above circuit may be coded as using VHDL shown in Figure 3.

```
1   -- input: spike(t3), 1_tau(t1), A(t0), G(t4)
2   -- output: H(t4)
3
4   M : decay_mem port map (clk=>clk, raddr=>A, dout=>B,
        we=>mem_we, waddr=>G, din=>H);
5   Mul : multiplier port map (clk=>clk, a=>A, b=>1_tau, c=>D);
6
7   process (clk)
8   begin
9       if ( clk'event and clk ='1') then
10          -- connect registers
11          C <= B;
12          E = C - D;
13          if (spike='1') then
14              H <= constant_1.0;
15          else
16              H <= E;
17          end if;
18      end if;
19  end process;
```

**Figure 3: VHDL code for designing example circuit**

In the code illustrated in Figure 3, each signal point or register in the circuit is identified by a unique signal name. Lines 4 and 5 of this code place the dual-port memory M and multiplier Mul in the circuit and specify the connections of the wire to each input and output port of the placed component. Lines 11–18 of this code specify the connections to the C, E, and H registers. The data flow of the above circuit diagram is described in the table shown in Figure 4.

**Figure 4: Data flow table**

In Figure 4, each row represents the data contents at various signal points in each clock cycle. To compute $N$ synapses, $N$ clock cycles constitute a single timestep, and the timestep repeats as the model time progresses. In this figure, the green cells indicate the flow of synapse 0, while the yellow cells represent the data flow of synapse 1. The result computed at H in a timestep is stored in the M memory, and subsequently, this value is read as the value of the signal point B in the next timestep to repeat the same computation. The output value of the $i^{th}$ synapse can be monitored by collecting the values of the $i^{th}$ H data from each timestep as shown on the right. As shown in this figure, all components continuously process data without interruption during the entire computation period. Memory such as M is embedded in the circuit, and the input and output ports are continuously read and written.

The throughput of such a fully pipelined circuit is always one data per clock cycle regardless of the complexity of the circuit. For example, when the system clock is 300 MHz, this circuit can compute 300 million synapses per second. It would be challenging to compute at such high speeds for a software program executing on a processor.

The described fully pipelined circuit is not new to this study and is a common method of designing digital circuits. However, the practical issue is to continuously provide input data to the circuit and continuously consume the output data computed by the circuit. From the system point of view, the key aspect is to connect the component pipeline circuits seamlessly and ensure that the entire system processes data continuously without stopping to maintain a high system throughput. The NM hardware architecture accomplishes this aspect using one computational node, while the proposed architecture uses multiple nodes achieving a linear speedup.

## 3 Hardware Design

### 3.1 System Configuration

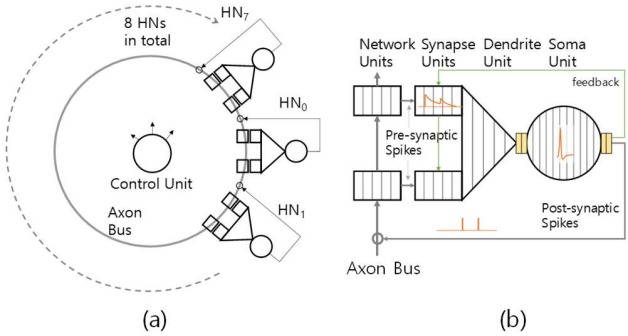

Figure 5: (a) System configuration, (b) block diagram of HN

One of the systems designed and implemented in this study can process 12 million neurons and consists of 8 HNs each computing 1.5 million neurons as shown in Figure 5(a). Each HN produces the spike information, i.e., binary data indicating whether an action potential is fired. The spike information of the neurons computed by each HN is shared with the other HNs through the axon bus. Using a synchronous communication scheme proposed in this study, each HN acquires the spike information of all neurons in the system immediately after the end of each network timestep.

### 3.2 HN

Each HN computes the neurons and synapses assigned to it in a time-division manner. The input of the HN is the spike information of all neurons computed in the previous timestep, while the output is the new spike information computed by this HN. One HN computes $P$ synapses and one HH action potential in every clock cycle.

As shown in Figure 5(b), each HN consists of $P$ network units (NUs), $P$ synapse units (SNUs), one dendrite unit (DU), and one soma unit (SU). As indicated by the name of each unit, the NU provides communication function between the neurons, SNU computes synaptic functions, DU adds up the synaptic currents, and SU computes the HH action potential. Here, the number $P$ of NUs and SNUs is a system design parameter, with $P = 2$ in the proposed design. Each unit of HN is designed as a fully pipelined circuit. In each HN, the NUs, SNUs, and DU operate with a network timestep of 1 ms, while the SU operates with a neuron timestep of 0.04 ms. The data of these two different timesteps are exchanged through memory buffers located at the outputs of the DU and SU. The data are cycled in the order of the NU, SNU, DU, SU, and NU. From the perspective of the computational architecture, an example of the neuronal and synaptic computation is provided here, and the hardware implementation of an arbitrarily complex neuron model can be developed in a similar manner.

### 3.3 Network Unit (reading part)

The NU converts the postsynaptic spikes computed from the SU into presynaptic spikes before sending it to the SNU, which corresponds to the communication between the neurons.

In all HNs, each NU and its corresponding SNU are in charge of $1/P$ of all synapses of all neurons computed by the HN. Here, the $a$th synapse of the $p$th NU-SNU pair of the $h$th HN can be defined as follows.

$$S_p^h(a) = \left\{ s_{i,j}^h \left| \begin{array}{l} i \mid \sum_{k=0}^{i-1}[N_k^h/P] \leq a < \sum_{k=0}^{i}[N_k^h/P] \\ j = \lfloor (a - \sum_{k=0}^{i-1}[N_k^h/P]) \times P \rfloor + p \end{array} \right. \right\}$$

Here, $s_{ij}^h$ is the $j$th synapse of the $i$th neuron processed by the $h$th HN, and $N_k^h$ is the number of input synapses of the $k$th neuron of the $h$th HN. This is equivalent to dividing the input synapses of each neuron into $P$-sized pieces and stacking them. For example, if $P = 2$ and the number of synapses of the first three neurons are 2, 5, and 4, respectively, the synapses will be arranged as shown in Figure 6.

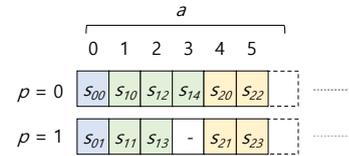

Figure 6: Synapse arrangement when $N_k^h = \{2, 5, 4, ...\}$

When the synapses of each neuron are divided into $P$-sized units, the remaining positions are filled with null synapses whose synaptic output value is always 0. The NU can be described separately as reading part and writing part. For convenience, only the reading part is described here, while the writing part will be described in a later section. The function of the reading part of each NU is to output the presynaptic neuron's spike information for each synapse processed by the NU, one data at each clock cycle. The output of the $p^{th}$ NU of the $h^{th}$ HN in the $t^{th}$ clock cycle after the start of a timestep can be defined as:

$$NU_p^h(t) = x_{m_{S_p^h(t)}}$$

Here, $x_j$ is the spike information of the $j^{th}$ neuron, and $m_S$ is the identification number of the presynaptic neuron of synapse $S$. In the reading part of the NU, the MM and MX memories are connected in series as shown in Figure 7. The memories are pipelined using the R1 and R2 registers.

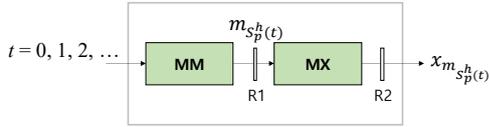

**Figure 7: Circuit of the NU reading part**

In this circuit, the topology information of the network is stored in the MM memory. In the $a^{th}$ address of the MM memory in the $p^{th}$ NU of the $h^{th}$ HN, the identification number of the presynaptic neuron of the $a^{th}$ synapse in the NU is stored and can be defined as follows.

$$MM_p^h(a) = m_{S_p^h(a)}$$

Here, $m_S$ is a 24-bit data that identifies one of all neurons in the system. Each MX memory in every HN has a memory space of $2^{24} \times 1$, and the spike information of all neurons in the entire system is stored. At the start of a new timestep, the address $t$ of the MM memory starts from 0 and increases by 1 at each clock cycle, which is provided by the control unit. Subsequently, the output of the NU appears in register R2 in the following clock cycles.

### 3.4 Synapse Unit

The SNU is responsible for computing the synaptic current that changes over time in response to spikes from presynaptic neurons.

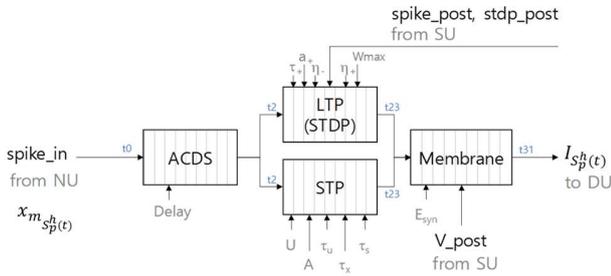

**Figure 8: Block diagram of the SNU**

In the proposed system, the SNU consists of the axonal conduction delay per synapse (ACDS), STP, LTP, and membrane parts as shown in Figure 8. The ACDS part supports spike delay from 0 ms to 24 ms. The STP, LTP, and membrane parts compute Equations 1, 2, 3, respectively

### 3.5 Dendrite Unit

The DU serves to sum the synaptic current of each neuron. The computation results are stored in the Netsum memory, which is a dual-port memory, where the write port operates as the network timestep and the read port as the neuron timestep.

### 3.6 Soma Unit

The SU computes the HH action potential. All HH equations are computed by 32-bit single-precision floating point operators. SU computes one action potential per clock cycle. In our high-capacity system, SU uses a total of 1.5 million clocks for one 40 μs neuron timestep, whereas NU, SNU, and DU use 37.5 million clock cycles to advance 1 ms network timestep. During one network timestep, the SU executes 25 neuron timesteps. Figure 9 shows the block diagram of the SU.

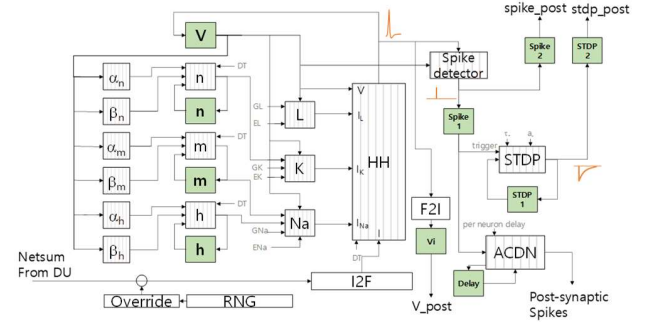

**Figure 9: Block diagram of the SU**

The external current from the DU is converted to a floating-point number before its use in the HH parts. The HH parts on the left side of Figure 11 contain four 32-bit state memories, namely, $V$, $n$, $m$, and $h$. The blocks with vertical stripes compute the HH action potential, as defined in Equation 4–7. Based on the action potential, the spike detector generates a spike when the action potential crosses a zero point from a negative value. The spike information stored in the Spike 1 memory is flushed out in the last timestep of every 25 neuron timesteps. The spike information is used to compute the postsynaptic STDP and per-neuron axonal conduction delay (ACDN), which has a delay range of 0–256 ms. The delayed spike is finally output as the postsynaptic spike of the SU.

### 3.7 Network Unit (writing part)

Each MX memory in the NU consists of 8 internal dual-port memories, as shown in Figure 10, with each dual-port memory occupying $2^{21}$ address spaces and the first 1.5 million address spaces implemented as a real memory.

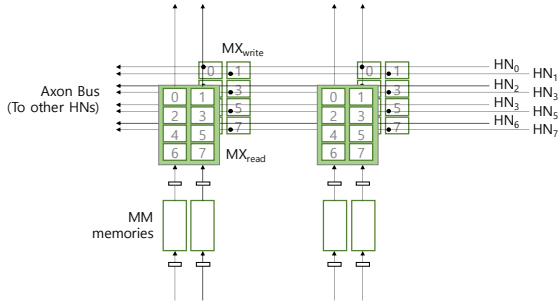

**Figure 10: Circuit of the NU writing part**

All the read address ports of these 8 memories are combined with a decoder circuit to form one MX memory with an address space of $2^{24}$. In contrast, the write ports of all 8 dual-port memories are used separately, and the dual-port memories located at the same location in all the 16 NUs of the system are tied together to form the axon bus. Therefore, the axon bus provides 8 write memory ports, and the $i^{th}$ port is connected to the spike output of the $i^{th}$ HN. When $i^{th}$ HN writes the output, the same contents are stored in the $i^{th}$ memories of all 16 NUs.

The control unit operates the network and the neuron timesteps with the timing shown in Figure 11.

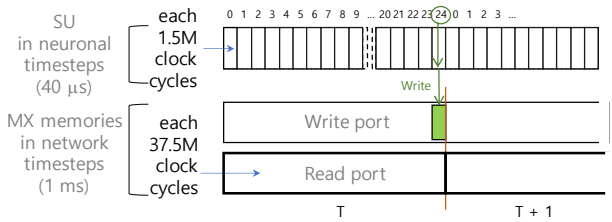

**Figure 11: Timestep sequence**

The spike data output by the SU at every 25th neuron timestep is written to the NU's MX memory via the axon bus. As the read and write ports of the dual-port memory access the same memory contents, the data written to the write port is immediately available from the read port. Therefore, following the completion of one network timestep, the new spike information of all neurons in the network is stored in the MX memory of all the Nus, which can be accessed directly from the reading part of the NU at the next timestep without delay. There is no delay in communication, and there are no restrictions on the network topology regardless of the physical location of the neurons.

## 3.8 Control Unit

**Generation of control signals**. The entire system is operated by the control unit, which transmits control signals of appropriate timing to each control point of the system. Most of the control signals are read and write addresses of the memories distributed over the system. As the memory addresses have different time references with the same sequence, the control unit can be easily implemented using a counter and shift register array.

**Attribute memories**. Instead of storing all the synaptic and neuronal attribute data in the memory, attribute set memories can be used to save memory space. For example, each SNU has a memory table of 1024 × 180 bits. By referring to the attribute set in this table, 180-bit rich attributes can be provided using just 10 bits of index data specific to each synapse. A similar method can be used for the SU.

**System initialization and logging**. The network topology and the initial state and attribute values of the synapses and neurons are set by separate back-end circuits, before system execution. In addition, checkpoints are captured as the system is running and the results are either displayed in real time through the HDMI port or recorded in the main memory and later accessed by the processor.

## 4 Implementation Results

### 4.1 Hardware Board and Chip

The proposed design was implemented on a Xilinx VU37P FPGA chip mounted on a VCU128 evaluation board as shown in Figure 12. The VU37P chip has substantial LUT and DSP resources to construct the circuit and supports on-chip memories that can construct up to 4,500 independent memories with a total capacity of 377 megabits. In particular, the chip has a built-in high bandwidth memory (HBM) that can support data rates of up to 460 GB/s with 8 GB capacity, therefore requiring no external memory.

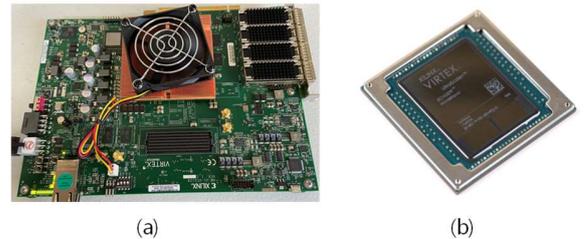

**Figure 12: (a) VCU128 board, and (b) VU37P FPGA chip**

The VCU128 board consists of a video output, an Ethernet interface, and a VU37P FPGA chip. A MicroBlaze soft processor and the neural network system designed in this work were implemented inside the FPGA. The Ethernet interface is connected to the user's PC. The PC houses a communication program and MySQL database. Following the design of a HH network by the user, the neuron, synapse, and attribute tables in the MySQL database can be configured using a database script program. The communication program reads the database and divides all the neurons into 8 even groups before sending this information through the Ethernet. Via the processor, the transmitted data are stored in the memories of each HN. The HH network is executed when the system execution command is issued from the PC, and real-time information of the pre-selected neurons and synapses is displayed on the monitor through the HDMI port as the timesteps progress.

### 4.2 Display Output

Figure 13 shows the display output of the proposed system. At the left of the figure, neuron states are displayed in a 1000 × 1000 area, where each pixel indicates the action potential of a neuron. The four areas on the right side of the screen show the internal statuses changing over time. From the top, the firing pattern of neurons, the state of the selected neuron, and the states of the two selected synapses are shown respectively. At the bottom right of the screen, the weight distribution is displayed.

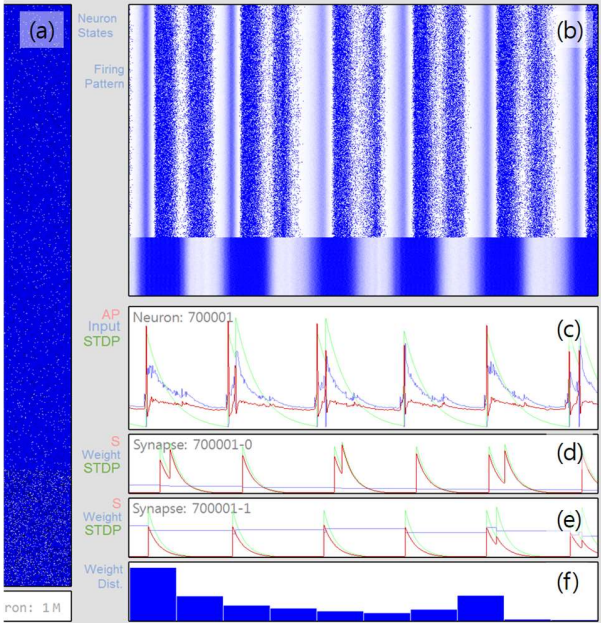

**Figure 13: Display output for a one-million neuron network: (a) neuron states, (b) firing pattern, (c) state of an excitatory neuron, (d) state of a spike-timing facilitation synapse, (e) state of a spike-timing depression synapse, and (f) weight distribution.**

### 4.3 Implementations

As the proposed design is of a synchronous system , in which all the data flow is predetermined, the system speed can be calculated in advance using the following equation.

$$\frac{model\_timesteps\_per\_s}{actual\_timesteps\_per\_s} = \frac{\frac{1}{timestep}}{\frac{freq\_ck \times N_{HN}}{N_{neuron}}}$$

Here, $freq\_ck$, $N_{HN}$, and $H_{neuron}$ are the clock frequency, the numbers of HNs, and neurons, respectively. Two versions of the design were implemented with different design parameters: high-capacity and high-speed implementations, each with a 300 MHz system clock. The speeds of the implementations are shown in Table 1.

**Table 1. Implementation Results of the Proposed Design**

| Neurons | Synapses | HNs | Timestep Network | Timestep Neuron | Time (s) / Model s | Implementation |
|---|---|---|---|---|---|---|
| 12 M | 600 M | 8 | 1 | 0.04 | **125.0** | High-capacity |
| 1 M | 50 M | 32 | 1 | 0.04 | **2.6** | High-speed |

The resource utilization of the high-capacity and high-speed implementations are shown in Table 2 and 3, respectively.

**Table 2. Resource Utilization: High-Capacity Implementation**

| Resource | NU | SNU | DU | SU | CU | Total used | Available on VU37P | Utilization |
|---|---|---|---|---|---|---|---|---|
| Onchip (Mbit) | 183 | 3 | 0 | 0.3 | 0 | 186 | 378 | 49.3% |
| Onchip rate (Gbit/s) | 10 | 864 | 0 | 614 | 0 | 1488 | - | - |
| HBM (MB) | 1717 | 4005 | 72 | 320 | 0 | 6114 | 8192 | 74.6% |
| HBM rate (GB/s) | 14 | 61 | 1 | 134 | 0 | 211 | 460 | 45.8% |
| LUT resource (k) | 1 | 13 | 4 | 118 | 18 | 154 | 1304 | 11.8% |
| DSP resource | 0 | 240 | 0 | 1061 | 6 | 1306 | 9024 | 14.5% |

In the high-capacity implementation, the network topology data (24 bits per synapse), synaptic (56 bits) and neuronal (192 bits per neuron) states are stored in the HBM memory and spike information in the MXs are stored in the on-chip memories.

**Table 3. Resource Utilization: High-Speed Implementation**

| Resource | NU | SNU | DU | SU | CU | Total used | Available on VU37P | Utilization |
|---|---|---|---|---|---|---|---|---|
| Onchip (Mbit) | 67 | 11 | 24 | 192 | 0 | 295 | 378 | 77.9% |
| Onchip rate (Gbit/s) | 38 | 3456 | 941 | 5530 | 0 | 9965 | - | - |
| HBM (MB) | 119 | 334 | 6 | 0 | 0 | 459 | 8192 | 5.6% |
| HBM rate (GB/s) | 48 | 245 | 0 | 0 | 0 | 293 | 460 | 63.7% |
| LUT resource (k) | 4 | 46 | 14 | 411 | 19 | 494 | 1304 | 37.9% |
| DSP resource | 0 | 838 | 0 | 3712 | 6 | 4556 | 9024 | 50.5% |

In the high-speed implementation, 32 HNs are used and the neuronal states in the SU are stored in the on-chip memories to speed-up the computation.

### 4.4 Performance Comparison

As listed in Table 4, two systems implemented in this work show noticeable results in speed and capacity among known HH systems.

**Table 4. Performance Comparison of HH Systems**

| Ref. | Platform | Model | Neurons | Synapses | Time (s) / Model second |
|---|---|---|---|---|---|
| [7] | GPU | HH[1] | 400 K | 3.2 M | 14400 |
| [16] | GPU | HH[2] | 1 M | 8 M | 400 |
| [17] | Multinode Phi KNL | HH[2] | 2 M | 2 B | 600 |
| [18] | Gordon Supercomputer | HH | 1 M | 4 B | 22 |
| **This** | FPGA | **HH** | **1 M** | **50 M** | **2.6** |
| **This** | FPGA | **HH** | **12 M** | **600 M** | **125** |

1) Extended to 11 ion currents, 2) Extended for and limited to the inferior olive of the brain

## 5 Discussion and Future Work

The proposed single-chip system demonstrated high speed mainly because all arithmetic operators are nearly 100% utilized during the entire computation cycle, with no complex overhead circuits such as sophisticated communication modules employed as in other systems [19][20]. This can be compared to a single brain chip with one million HH physical neurons, each computing with floating-point precision and running in a near real-time. The proposed system has the same functionality as the above system, which, in addition, supports unlimited network topology.

In the proposed implementations, many computational resources (LUTs and DSPs) from the FPGA chip remain unutilized, as shown in Table 2. These resources can be used to implement complex multi-compartmental models [20] in the future using fully pipelined circuits, as shown in Figure 14.

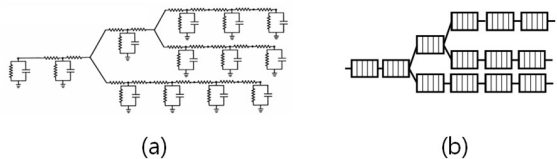

**Figure 14: (a) Concept diagram and (b) pipelined circuit design of a multi-compartmental model**

It should be noted that a larger system can be built using multiple FPGA chips. In this case, only NUs need to be changed to incorporate a larger number of neurons. For example, 100 FPGA chips each computing 12 million neurons can be combined to implement a system supporting 1.2 billion HH neurons and 60 billion synapses. In this case, each MX memory in the NU would require on-chip memories with 1.2 gigabits capacity, which current FPGA technology cannot support. The design of a separate memory chip dedicated to the NUs could provide a potential solution.

The object of this study was to implement a high-performance computing project. With respect to developing tools for brain research, the requirements from neuroscientists were not sufficiently reflected. With the growth of neural networks, a network authoring tool, which current work lacks, is necessary. Ultimately, the authoring system that automatically creates a hardware design from the equations of the neuron model could be developed. The input and output interfaces need to be further optimized for the artificial brain system to support AI research. For example, the pixel values of the deep learning training images can be mapped to the input neurons and dynamically changed.

## 6 Conclusion

In this paper, a hardware architecture suitable for the artificial brain implementation of a large network of complex neuron models was demonstrated. This new hardware architecture can be used for developing artificial brains, which can, in turn, be used for further studies of the brain, implementation of neural prosthetic devices, and development of new machine learning algorithms. Further study is required for the multi-chip systems that compute large-scale networks of multi-compartmental neurons.